# Simultaneous ambient pressure X-ray photoelectron spectroscopy and grazing incidence X-ray scattering in gas environments

*Heath Kersell[1], Pengyuan Chen[2], Henrique Martins,[1,3] Qiyang Lu[1,4], Felix Brausse,[5] Bo-Hong Liu,[1,5] Monika Blum,[1,5] Sujoy Roy,[1] Bruce Rude, [1] Arthur Kilcoyne,[1] Hendrik Bluhm,[1,5,6] Slavomír Nemšák[1]*

[1]Advanced Light Source, Lawrence Berkeley National Laboratory, Berkeley, California, 94720, United States

[2]Department of Electrical Engineering and Computer Sciences, University of California, Berkeley, California, 94720, United States

[3]Department of Physics, University of California, Davis, California, 95616, United States

[4]Department of Materials Science and Engineering, Stanford University, Stanford, California, 94305, United States

[5]Chemical Science Division, Lawrence Berkeley National Laboratory, Berkeley, California, 94720, United States

[6]Department of Inorganic Chemistry, Fritz Haber Institute of the Max Planck Society, 14195, Berlin, Germany

*Keywords: Ambient Pressure, X-ray Photoelectron Spectroscopy, Grazing Incidence X-ray Scattering, Silver Behenate*

**ABSTRACT:** We have developed an experimental system to *simultaneously* observe surface structure, morphology, composition, chemical state, and chemical activity for samples in gas phase environments. This is accomplished by simultaneously measuring X-ray photoelectron spectroscopy (XPS) and grazing incidence X-ray scattering (GIXS) in gas pressures as high as the multi-Torr regime, while also recording mass spectrometry. Scattering patterns reflect near-surface sample structures from the nano- to the meso-scale. The grazing incidence geometry provides tunable depth sensitivity while scattered X-rays are detected across a broad range of angles using a newly designed pivoting-UHV-manipulator for detector positioning. At the same time, XPS and mass spectrometry can be measured, all from the same sample spot and in ambient conditions. To demonstrate the capabilities of this system, we measured the chemical state, composition, and structure of Ag-behenate on a Si(001) wafer in vacuum and in $O_2$ atmosphere at various temperatures. These simultaneous structural, chemical, and gas phase product probes enable detailed insights into the interplay between structure and chemical state for samples in gas phase environments. The compact size of our pivoting-UHV-manipulator makes it possible to retrofit this technique into existing spectroscopic instruments installed at synchrotron beamlines. Because many synchrotron facilities are planning or undergoing upgrades to diffraction limited storage rings with transversely coherent beams, a newly emerging set of coherent X-ray scattering experiments can greatly benefit from the concepts we present here.

## 1. Introduction

Nano- and atomic- scale understanding of chemical and structural processes is highly sought-after in many technologically relevant areas; from photovoltaics [1] and information storage [2] to batteries [3] and heterogeneous catalysis. [4] Additionally, practically all important technological or natural processes take place in ambient gas environments. These environments often induce complex relationships between structure, chemistry, and function. For example, as dimensions reach the nanoscale, transition metals and transition metal oxides exhibit size dependent behavior, contributing to heavy metal migration in geological systems. [5] [6] [7] Reactant gas environments can lead to pressure-dependent adsorbate coverages [8] [9] [10], and potentially initiate nanocluster formation at catalyst surfaces, subsequently altering reaction pathways. [8] [11] [12] Active species in bimetallic alloy catalysts can segregate under reaction conditions, altering their *operando* surface composition. [13] [14] Polyamide membranes for water purification exhibit correlations between water permeability and the presence of nanoscale Turing structures. [15] Detailed understanding of the correlation between chemistry, structure, and function in these systems requires multimodal investigations. Additionally, each of these processes occurs in heterogeneous environments, and measuring them in ambient gas mixtures requires in-situ or operando techniques.

In the past two decades, a number of operando spectroscopic and structural probes were developed to study heterogeneous interfaces. [16] [17] For example, environmental transmission electron microscopy (ETEM) enables cross-sectional elemental mapping in ambient pressure environments. [18] Polarization-modulation infrared reflection-absorption spectroscopy (PM-IRRAS) provides an operando probe of interfacial adsorbates. [16] Dynamic surface restructuring at pressures from ultrahigh vacuum (UHV) to the bar regime can be measured with atomic resolution using high pressure scanning tunneling microscopy (HPSTM) and atomic force microscopy (HPAFM). [17] [19] [20] Operando studies of surface chemical composition and chemical states, as well as adsorbate species can be pursued with ambient pressure X-ray photoelectron spectroscopy (APXPS). [21]

In spite of the growth of operando methods over the last two decades, understanding the detailed relationship between dynamic surface restructuring and chemical transformations remains a difficult task. One challenge lies in correlating chemical transformations with structural evolution, each measured with separate techniques. Currently, this is often done using separate samples, separate measurement sets, and frequently in entirely separate experimental systems. Differences in pressure measurements, sample temperatures, time resolutions, and sample quality can lead to difficulties comparing structural or chemical changes in separate measurements. This opens a gap between the two experiments, complicating the interpretation of separately collected data, and hindering the accuracy of such separate experiments. Therefore, multi-modal approaches, like the one presented in this work, have been undertaken at various synchrotron facilities, with a few successful results. For example, a combined APXPS and IRRAS study utilized the sensitivity introduced by IRRAS to identify species that were not detected in APXPS measurements. [22] Some experimental systems are now also equipped to measure X-ray based spectroscopies and X-ray diffraction (XRD). [23] For example, a setup at Argonne National Laboratory allows for simultaneous analysis of chemistry and structure using hard X-ray photoelectron spectroscopy and X-ray diffraction in a specifically designed pulsed laser deposition chamber. [24] To address the need for simultaneous structural and chemical probes under ambient conditions, we have combined APXPS with grazing incidence X-ray scattering (GIXS). This combination enables the *simultaneous* correlation of sample chemical states and adsorbate species, with surface structures that emerge in heterogeneous environments, where ambient pressures reach into the Torr regime.

## 2. Combining ambient pressure photoemission and X-ray scattering

In XPS, X-ray photons incident on a material cause photoelectron emission if the photon energy exceeds the binding energy of electrons in the material. The resulting photoelectron intensity at various kinetic energies provides information about surface chemistry, composition, valence states, and adsorbate coverage. [25] Because electrons are strongly scattered by matter, these photoelectrons are predominantly from the surface and near-surface regions, with typical probing depths in the few nm range. Electrons are also strongly scattered by gas molecules, and thus most XPS experiments are performed in high or ultra-high vacuum conditions. To measure in ambient pressures, APXPS apparatuses are equipped with a small (sub-mm) aperture in front of the sample surface, which photoelectrons pass through. [21] [26] This aperture leads to a differentially pumped electrostatic lens system, and then to a hemispherical electron energy analyzer, as shown in Fig. 1. [27] The differential pumping strongly reduces scattering of photoelectrons in the lens system by gas molecules, allowing them to reach the hemispherical analyzer where differential pumping maintains high vacuum.

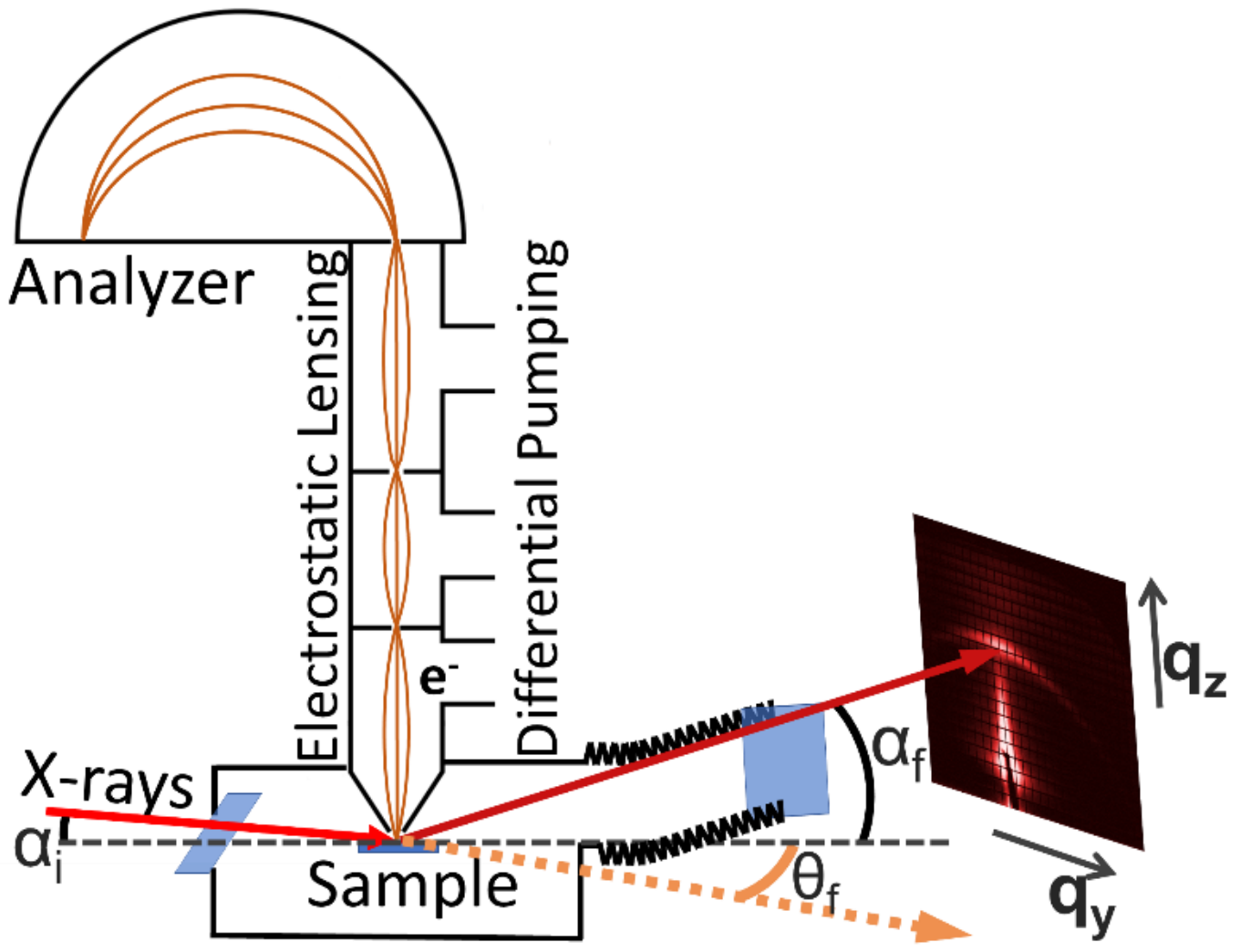

Fig. 1. Schematic of a combined ambient pressure XPS/GIXS system. Incident X-rays pass through a 100 nm thick silicon nitride window (blue, lower left side) separating the beamline from ambient gas environments in the sample chamber. Photoelectrons are collected by the differentially-pumped electrostatic lens system and analyzer, which are separated from the sample environment by a small aperture next to the sample. In the APGIXS measurements, photons scattered from the surface are measured by a detector (right side). A flexible edge-welded bellows (zig-zag lines) mounted on a rotating 2D joint allows the detector to be scanned across scattering angles of interest. The X-ray detector is separated from the gas environment by an X-ray transparent silicon nitride window with a 150 nm thickness (blue, lower right side). The two silicon nitride windows (blue) and the cone aperture separate the beamline, X-ray detector, and the XPS analyzer from the gas phase environment present in the sample chamber. X-ray scattering angles correspond uniquely to specific values of $q_y$ and $q_z$, where the longitudinal and transverse scattering angles are denoted $\alpha_f$ and $\theta_f$, respectively. The detector range is $0° \leq \alpha_f \leq 20°$, and $-20° \leq \theta_f \leq 20°$.

While some photons incident on the sample are absorbed to produce photoelectrons, some are also elastically scattered. The resulting scattering pattern reflects the arrangement of scattering centers in the sample (i.e. the sample structure). Using X-rays incident at grazing angles enables X-ray scattering measurements in the reflection geometry (as in Fig. 1), [28] (opposite to the transmission geometry) allowing simultaneous measurements of ambient pressure GIXS (APGIXS) and APXPS.

Scattered X-rays from samples form an interference pattern with maxima (constructive interference) at angles inversely related to the location of the scattering centers. When the scattering patterns are generated by X-rays at grazing incidence, the corresponding X-ray scattering techniques are broadly grouped into grazing incidence wide angle X-ray scattering (GIWAXS) and grazing incidence small angle X-ray scattering (GISAXS).

Depending on the wavelength of the incident X-rays, GIWAXS provides access to atomic scale information at large scattering angles, making it ideal for studies of crystal structures and their transformations. [30] [31] Meanwhile, at smaller scattering angles, GISAXS has been applied to the study of more extended structures in thin film growth, nanoparticle stability, block copolymer thin film evolution, and quantum dots, for example. [32] [33] [34] [35] There is no unambiguous delineation between the small and wide-angle regimes in GIXS, and combined GISAXS/GIWAXS investigations provide a detailed picture of crystal structures and the growth or evolution of structures from the nanoscale to the mesoscale. [36]

## 3. Description of the experimental apparatus

Measurement of GIXS across the small- and wide-angle regimes in ambient gas environments poses a number of challenges. For example, simultaneous APGIXS and APXPS measurements require (i) that X-rays be incident on samples at positions that optimize the photoelectron intensity reaching the analyzer, while also scattering into angles accessible by the X-ray detector. Detecting structures from the atomic- to the meso-scale also requires a detector capable of both (ii) measurement over a wide range of scattering angles, and (iii) high angular resolution (especially for coherent scattering and observing speckles). Additionally, (iv) the photon detector itself often needs to be in vacuum, isolated from gas phase sample environments.

To address these issues, we developed the experimental configuration illustrated in Fig. 1. In this layout, a custom UHV sample chamber is connected to the X-ray source at beamline 11.0.2 of the Advanced Light Source in Berkeley [37] with a practical photon energy range of 180 – 1800 eV, and to a differentially pumped commercial APXPS analyzer (SPECS Phoibos 150 NAP), as well as a moveable differentially pumped X-ray detector (Andor iKon-L SO CCD). X-ray transparent silicon nitride windows separate the sample environment from both the X-ray source (beamline) and the X-ray detector. The APXPS detector is separated from the sample environment by the differential pumping stages as discussed in the previous section. Although not shown in Fig. 1, the chamber is also connected to a sample preparation and load-lock chamber, and a four-axis sample manipulator. How this layout addresses the challenges (i-iv) enumerated in the previous paragraph is discussed next.

*Addressing challenge (i):* A commercially available four-axis manipulator (Thermionics Inc.) enables three dimensional cartesian positioning of the sample, and polar rotation with respect to both the X-ray beam source and the APXPS analyzer. This positioning system enables sample alignment such that grazing incidence X-rays generate photoelectrons which enter the APXPS aperture, while some photons simultaneously scatter from the sample toward the detector. In our system, a ~100 μm wide X-ray beam is incident on the sample surface approximately 0.5 mm from the 0.3 mm wide APXPS aperture. When the X-ray beam is at grazing incidence with respect to the sample surface, this geometry elongates the X-ray beam along the beam direction, generating scattered X-rays from an elongated sample region.

The grazing incidence geometry also enables control over the information depth for these techniques: the penetration depth of X-rays drastically decreases with decreasing incidence around the critical angle, which can be as high as several degrees at 1000 eV. Meanwhile the emission angle in XPS affects the depth contribution of photoelectrons reaching the analyzer. We note that in principle, a transmission scattering geometry could also be used, although sample thickness becomes an important consideration in that case.

*Addressing challenge (ii):* To measure GIXS over a wide range of scattering angles, we developed a new pivoting-UHV-manipulator capable of scanning an X-ray detector in two angular dimensions (Fig. 2a). As shown in Fig. 2, this device is attached to our vacuum chamber by a 6” CF flange at one end (left side of Fig. 2a). At the opposite end is a 6” CF flange where a vacuum compatible X-ray detector can be mounted (right side of Fig. 2a). UHV is maintained between these two sides by a combination of flexible edge welded bellows, 4” stainless steel tubes, and 6” CF flanges. At the chamber-side, the pivoting-UHV-manipulator contains a pair of rotation axes that enable movement of the detector in two angular dimensions (straight blue arrows in Fig. 2a). The vacuum bellows is flexible, bending and maintaining vacuum when the detector is moved. Detector motion is controlled by a pair of precision linear actuators, with a step resolution of 10 μm. The top of each actuator is attached to the pivoting-UHV-manipulator by a rotatable ball joint, so that actuator extensions or retractions precisely translate the manipulator across two angular directions (Fig 2a). Each actuator base is connected to a universal joint fixed to the system frame (not shown), allowing the actuators to pivot while controlling the detector position. This layout facilitates movement of the detector in two angular directions across a pseudo-spherical cap, illustrated by green curved arrows at the right side of Fig. 2a. An important aspect of X-ray scattering measurements is the range of available scattering angles where the camera has direct, unobstructed line-of-sight to the sample. This range is primarily determined by the size and position of the main vacuum flange between the chamber and pivoting-UHV-manipulator. This set of unobstructed angles determines the range over which scattered X-rays can be detected, and is illustrated in Fig. 1 by the red arrow scattered by the sample. We optimized this range, and the range of detector motion, to coincide with X-ray scattering angles relevant to many ambient pressure investigations. Scattered X-rays beyond the angular range represented by the red arrow in Fig. 1 will be obstructed. In our present layout, the angular range can be as large as approximately ±12° azimuthally and 0° to +24° in polar direction (with respect to the sample), which accommodates measurement of features as small as approximately 2 nm for the photon energy range available at beamline 11.0.2 at the ALS. This range can be increased by using a larger diameter bellows and a larger port on the chamber, which provides more space for scattered X-rays to pass unobstructed to the detector. This design facilitates GIXS measurements in both the small-angle and wide-angle regimes. One major advantage of this design is its relatively compact size. As shown in Fig. 2b, the entire pivoting-UHV-manipulator, which moves the X-ray detector, is not much larger than our eight-inch-wide UHV sample chamber. This is significantly smaller than conventional SAXS flight tubes, by approximately an order of magnitude or more. As a result, this design can be implemented for GIXS measurements in existing systems whose size is constrained by the amount of experiment floor available around the instrument.

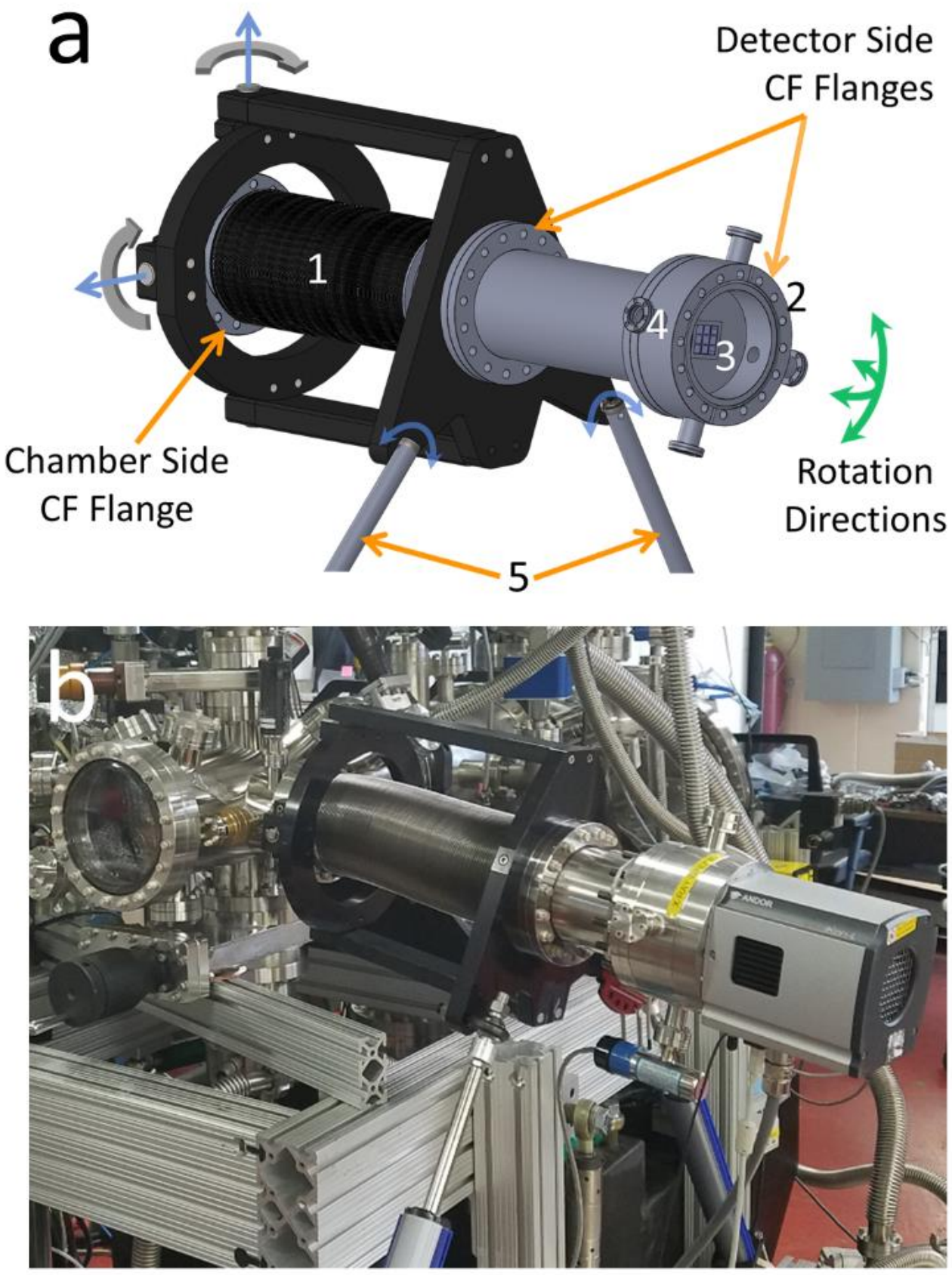

Fig. 2. Pivoting-UHV-manipulator. The model (a) and picture (b), show an edge welded bellows (1) attached to two CF flanges. The flange indicated at the left (chamber side) is rigidly attached to a UHV chamber. The flange at the right (detector side) pivots around a set of bearings (at the chamber side), allowing it to independently rotate about the two bearing axes (designated by straight blue arrows). This facilitates movement of a detector (a CCD camera in our case) through a solid angle (green curved arrows) with approximately ±12° around either axis. The detector mounts onto the CF flange at the right (2), where it is separated from the gas environment by an X-ray transparent 150 nm thick silicon nitride window (3). The area between the window and detector can be pumped through one of the CF mini ports (4). Control over the rotation is facilitated by the use of a pair of approximately orthogonal precision linear actuators (5).

*Addressing challenge (iii):* One critical experiment that requires high angular resolution is coherent X-ray scattering. Coherent illumination of a distribution of scattering centers gives rise to a so-called 'speckle pattern', which is characterized by interference of photons scattered by that distribution. Changes in the speckle pattern can be measured to understand changes in the distribution of scattering centers. To extract meaningful information about the evolution of a distribution of scattering centers in such measurements, a detector must have resolution at least as fine as the speckle size, $\sigma$, which depends on the photon wavelength, $\lambda$, the distance between the sample and detector, *D*, and the width of the coherently illuminated area, $a$, as [38]

$$\sigma = \frac{\lambda D}{a}. \tag{1}$$

In the experimental setup described in this section, we use a CCD camera as our detector, with 2048 × 2048 pixels and 13.5μm × 13.5μm pixel size (Andor iKon-L SO). Given formula (2), the minimum sample-detector distance for a 30 μm spot-size and 1.5 keV radiation at beamline 11.0.2.1 of the Advanced Light Source is ca. 50 cm. To measure with high angular resolution across the small- and wide-angle regimes, and hence probe features from approximately the single nanometer scale to the mesoscale, the pixel size of the detector and its positioning have to work with a resolution on the order of ~10 μm. Such conditions are also necessary to pursue more sophisticated investigations of, for example, structural dynamics.

*Addressing challenge (iv):* Using a CCD camera for measuring scattered X-rays presents a unique challenge in ambient pressure environments. To lower electronic thermal noise, CCD detectors are usually cooled using multiple-stage Peltier elements. As a result, any humidity or other condensable gases can coat the camera surface, causing electrical shorts and artifacts in the imaging, and inevitably reducing the lifetime of the camera. It is therefore necessary to isolate the detector from the gas phase sample environments. For the experimental setup we have described in this section, this is accomplished using a 150 nm thick X-ray transparent silicon nitride membrane with an area of 27 mm × 27 mm (Silson Ltd). The window is divided into a 3×3 array of 8.33 mm sections, divided by 1 mm thick frames to enhance the membrane stability against high pressure differentials (Fig. 2a). Scattered X-rays pass through the silicon nitride membrane to reach the CCD detector, while the membrane isolates the CCD chip from the gas phase environment of the sample chamber. These membranes are stable up to pressure differentials of several tens of Torr, which enables X-ray scattering measurements in ambient pressure gas environments across a similar pressure regime to that accessible in our APXPS system (<20 Torr). This has the added benefit of protecting the CCD chip from corrosive gases, enabling the study of a number of additional systems in, e.g., corrosion science. A set of four 1.33" CF flanged ports on the camera side of the X-ray window (shown in Fig. 2) enables the use of differential pumping, pressure measurement, and other diagnostic devices (e.g. a photodiode and beamstop).

To monitor the gas composition, including the formation of reaction products, we also installed a mass spectrometer in the first differential pumping stage of the electrostatic lensing system (Fig. 1), where the pressure is about four orders of magnitude lower than in the sample chamber. This enables mass spectrometry measurements to accompany the aforementioned APXPS and APGIXS. Finally, *in-situ* sample preparation is enabled by a sample heater and thermocouple on the sample manipulator, and by an ion gun and gas leak valves in the preparation chamber.

## 4. Proof-of-principle measurements of Ag-behenate

To confirm the capabilities of this newly built system, we measured changes in Ag-behenate ($AgC_{22}H_{43}O_2$) layers deposited on $SiO_x$ as a function of temperature in the presence of $O_2$ gas. The Ag-behenate molecule (Fig. 3a) contains a long-chain silver carboxylate which, when deposited onto a substrate, forms dimers that arrange into a layer-by-layer structure with an interlayer spacing of approximately 5.84 nm. Because of its regular interlayer spacing, Ag-behenate is a common reference standard in X-ray scattering experiments. [39] [40] Its layered structure yields a scattering pattern characterized by rings of high intensity at $Q$ = 0.108 Å$^{-1}$ and at successively higher order scattering maxima. [41] In our measurements, Ag-behenate was deposited onto a Si(001) surface (covered by a native oxide layer) by drop-casting a Ag-behenate-in-ethanol solution onto the surface and allowing the solution to dry in air. Once placed in vacuum, this system generated a clear scattering feature at 0.108 Å$^{-1}$ in our GIXS measurements, shown in Figs. 3b and c. In Fig. 3b, we note that the intensity of the scattering feature decreases somewhat from the side of the image where the recording was started (right) to the end (left), suggesting that over the course of the image, some beam damage may occur.

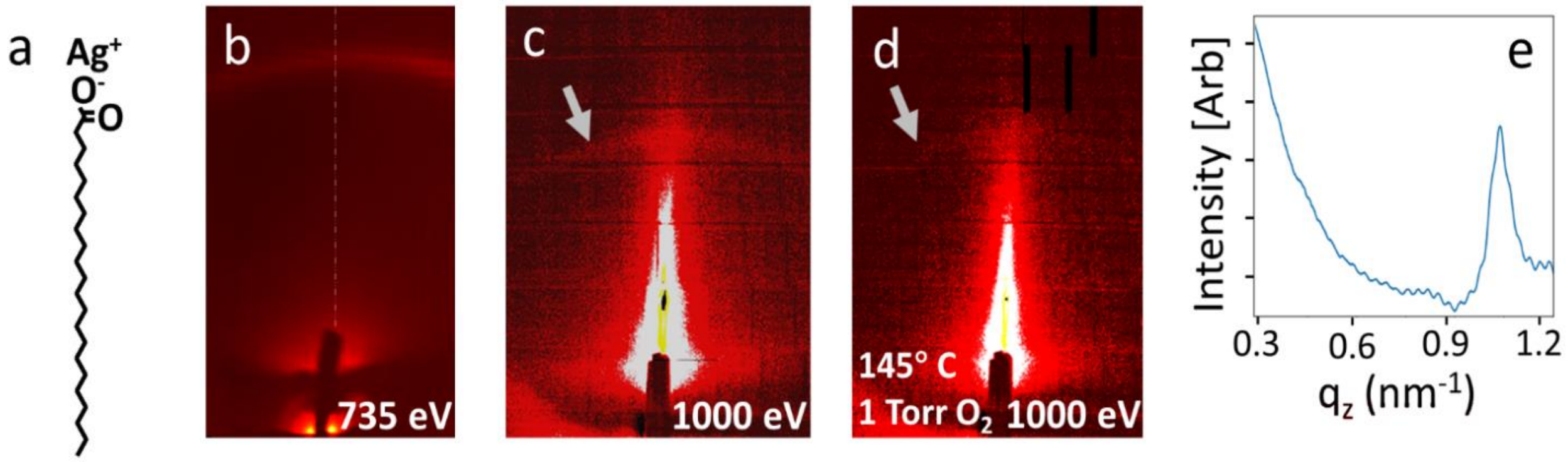

Fig. 3. X-ray scattering from Ag-behenate. a) Ag-behenate molecular structure; b) scattering pattern containing the first order peak formed by X-rays scattered from Ag-behenate. c) Low exposure acquisition (117 seconds) of Ag-behenate scattering pattern. A grey arrow designates the scattering intensity maximum from the Ag-behenate layers. d) Low exposure acquisition of Ag-behenate scattering pattern after heating in oxygen gas. The grey arrow designates the location where the scattering feature from Ag-behenate layers existed before heating. e) Intensity line profile across the vertical centerline shown in the scattering pattern in b). Note that 735 eV photons were used in (b), and 1000 eV photons were used to produce (c) and (d). The different scattering angles for (b) as compared with (c) and (d) are generated by the different photon energy used to generate the scattering pattern.

As a result, we recorded the images shown in Fig. 3c,d, and the spectra shown in Fig. 4, on a different sample, and with less exposure to the X-ray beam to mitigate beam damage. Figs. 3c,d were recorded with exposure times of 117 seconds per image. Because the photon flux at beamline 11.0.2 of the ALS is $3.3\times10^{12}\ \frac{\text{photons}}{\text{mm}^2\cdot\text{s}}$, that duration yields approximately $1.5\times10^{14}$ photons incident on the sample per image, more than an order of magnitude lower than for the image in Fig. 3a. The APXPS spectra discussed next were also collected in the same measurement set, increasing the total exposure to $3.5\times10^{14}$ photons per condition, still significantly less than for the image shown in Fig. 3a.

Fig. 4a shows C 1s XPS spectra for Ag-behenate/Si(001) under various temperatures and oxygen pressures. Ag-behenate contains a $CH_x$ backbone with 21 carbon atoms and a terminating carboxyl group (-COO) with two O atoms. As a result, the XPS intensity in the C 1s region is expected to have its largest contribution from C-C species, generally appearing near 284 eV. [42] Adventitious carbon species accumulate on surfaces exposed to air, and also contribute to the C 1s spectrum near 284 eV. [43] The bottom spectrum in Fig. 4a was taken in vacuum after introducing the sample into the chamber, with overlapping C-C and adventitious carbon peaks appearing in the C 1s region, along with a peak produced by the carboxyl group in Ag-behenate at 286.9 eV. [44] [45] The ratio of these two peaks was approximately 10:1 (Fig. 4a). This contrasts with the composition of the molecule, which has 22 C atoms, only one of which exists in a carboxyl group. Experimentally, the ratio of these two peaks reflects not only the ratio of each type of carbon species in the molecule, but is also influenced by the presence of contaminant species on the sample, by the orientation of the molecules on the surface, and by the integrity of the molecular structure. Regarding the molecular integrity, Ag-behenate is known to decompose upon prolonged exposure to air and light, which can contribute to different spectral weights in the C 1s spectra, and create a diffuse photon scattering background. Hence, the vacuum XPS in the C 1s region in Fig. 4a reflects that Ag-behenate has undergone some decomposition, and the sample contains other carbon-containing species, potentially altering the area ratios of the relevant peaks. Fig. 4b shows the Ag 3d spectra corresponding to each condition from Fig. 4a. In vacuum, the Ag 3d spectrum contains a peak at 367.8 eV, characteristic of $Ag^+$. [46] The presence of a feature above 370 eV is indicative of Ag satellite features which can contain contributions from $Ag^0$ and/or Ag partially oxidized by interaction Ag-behenate. [47]

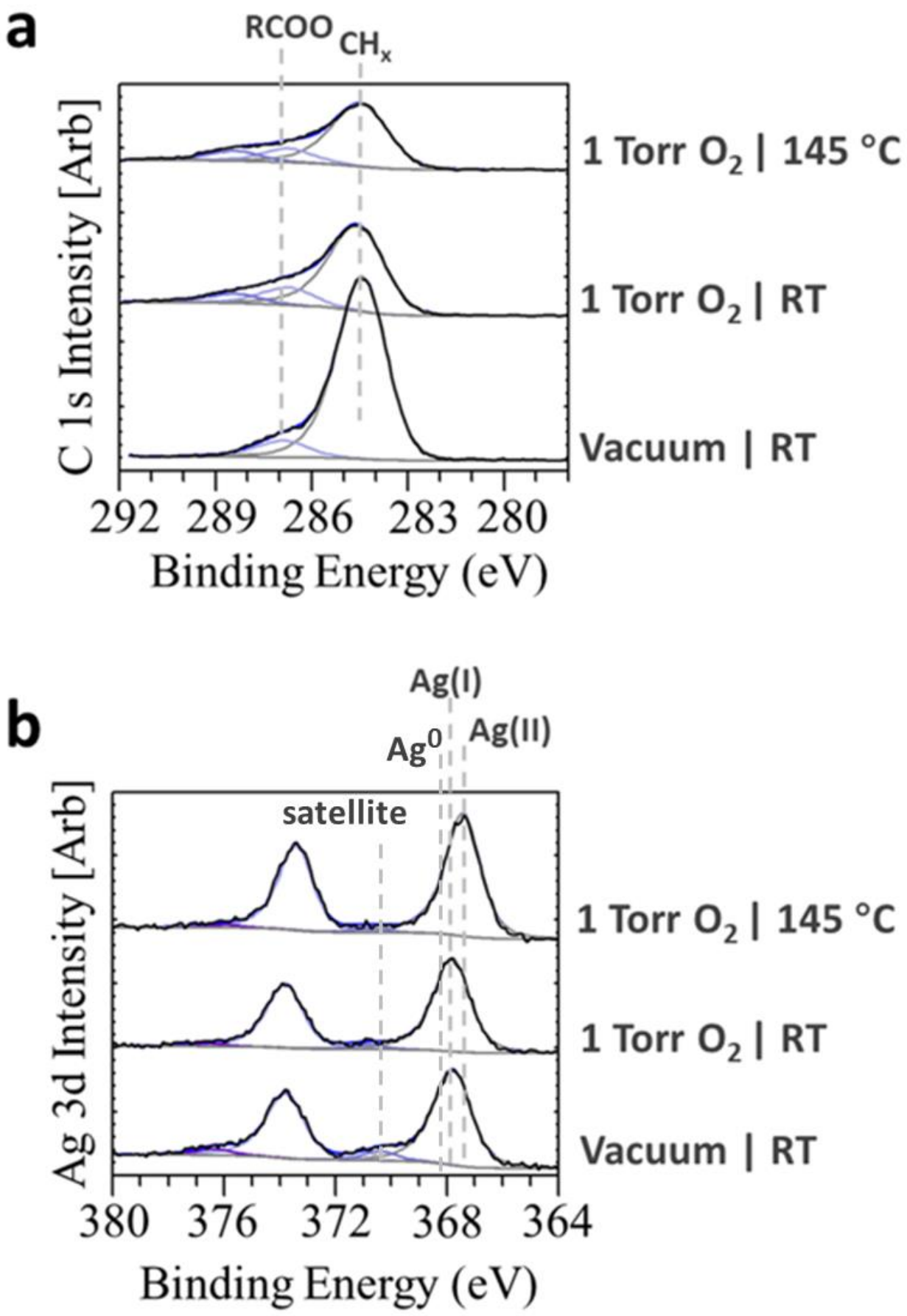

Fig. 4. (a) APXPS of the C 1s region on Ag-behenate under various conditions. Three features appear in the spectrum at 283.8 eV, 285.9 eV, and 287 eV. The C-C peak near 284 eV contains contributions from the hydrocarbon chain in Ag-behenate and from adventitious carbon. The RCOO peak arises from the carboxyl group in Ag-behenate. The appearance of the higher BE peak is characteristic of more highly oxidized carbon species in response to added oxygen. (b) Ag 3d region. Contributions appear from $Ag^+$ in Ag-behenate, some metallic Ag prior to adding oxygen, and $Ag^{2+}$ upon heating in $O_2$(g). Each spectrum in (a) and (b) is fitted with a Shirley background shape, represented by a grey line at the base of each spectrum. Photons with 1000 eV excitation energy were used to collect the C 1s and Ag 3d spectra.

After characterizing the surface in vacuum, we added 1 Torr of $O_2$ gas into the chamber. Oxygen exposure reduced the intensity of the C-C peak at 284.5 eV, suggesting a reaction with some of the carbon species at the surface. At the same time, the intensity of the peak at 286.9 eV remained approximately the same as in vacuum. This indicates that the carboxyl groups in Ag-behenate do not rapidly decompose upon $O_2$ exposure in these conditions, even when parts of the molecule decompose slowly when exposed to light and ambient air conditions. The decrease in the C-C peak is accompanied by the appearance of a feature at 288.5 eV, characteristic of more highly oxidized carbon species resulting from exposure to $O_2$ gas. [48] In the C 1s region, when the sample was heated to 145 °C in 1 Torr of $O_2$, only a 10% decrease in the intensity of the C-C peak was observed, indicating that the remaining carbon did not react readily with oxygen at this temperature. Nevertheless, the scattering feature generated by regularly spaced layers of Ag-behenate nearly vanished. Ag-behenate undergoes thermally induced carbon chain folding above 120 °C, where it adopts a new molecular packing arrangement that decreases the Ag-behenate interlayer spacing. [40] This is consistent with our observation that the feature due to scattering from regularly spaced Ag-behenate layers nearly vanishes under these conditions.

The chemical transformation of Ag-behenate, as observed by APXPS, is consistent with the results of the simultaneous scattering experiments. Nevertheless, the altered packing arrangement of the Ag-behenate layers has a structure which should also produce a scattering pattern detectable in the angular range of our GIXS system [40]. However, we note that the C-C and carboxyl intensities decreased somewhat while heating in oxygen, and at the same time, the BE of the Ag 3d peak shifted to 367.3 eV upon heating in $O_2$. This latter BE shift is characteristic of a transformation to $Ag^{2+}$ [46], suggesting that the Ag atoms from the Ag-behenate layered structure further oxidize into AgO. The oxidation of Ag under these conditions could coincide with a change in the structure of the Ag-behenate layers, explaining the disappearance of the diffraction ring. Additionally, the diminished intensity of both the carboxyl and C-C peaks suggests that the molecules begin to decompose with some components becoming volatile.

## 5. Conclusions

These combined APXPS and APGIXS measurements on Ag behenate illustrate the power of correlating chemical and structural transformations for samples in gas phase environments. Changes in the Ag behenate layered structure were observed using APGIXS, and were correlated with changes in the Ag oxidation state and carbon chemistry observed in the Ag 3d and C 1s XPS regions. Combining APXPS and APGIXS provided a simultaneous picture of the chemical state and structure of the sample, and provided a detailed picture of their evolution in an $O_2$ gas environment and in different thermal conditions. Our approach to understanding correlations between structure and behavior is applicable in a variety of fields, where important phenomena often involve complex relationships between structure, chemistry, and activity. [5-15]

The geometry of APXPS measurements involves X-rays incident on surfaces in a way that inherently enables scattering measurements in the reflection geometry (Fig. 1). Additionally, our development of a compact pivoting-UHV-manipulator to control the detector motion makes this configuration highly customizable. Using an X-ray transparent $SiN_x$ membrane to isolate the X-ray detector enables GIXS measurements for samples in a wide variety of gas phase environments. These developments make it possible to retrofit many existing APXPS, XPS, and other experimental systems to include simultaneous (AP)-GIXS measurements with relatively low cost and effort.

Further related techniques can also be enabled using this approach. For example, X-ray photon correlation spectroscopy (XPCS) measures correlations in scattering intensities from a distribution of scattering centers illuminated by coherent photons. The evolution of the correlation between speckle patterns over time provides a measurement of the evolution of scattering center distributions in a sample. [49] This opens new possibilities to study dynamic systems and transformations on their natural timescales. Several soon-to-be-emerging diffraction limited synchrotron sources around the world will offer coherent flux intensities increased by orders of magnitude, allowing measurements with time resolution into the microsecond and nanosecond regimes.

## AUTHOR'S CONTRIBUTIONS

H.B., S.N. initiated the project. H.K., S.R., A.K., S.N. contributed to the system design. H.K., S.N. performed the experiments with support from B.H.L. and M.B. Data analysis was performed by H.K. with support from Q.L., H.M., P.C., and F.B. H.K. and S.N. wrote the manuscript, S.N. supervised the project.

## ACKNOWLEDGEMENTS

The instrumentation presented in this work was developed through the Laboratory Directed Research and Development (LDRD) Program of Lawrence Berkeley National Laboratory under a grant titled "Correlation of structural and chemical processes at interfaces under operating conditions using multimodal ambient pressure X-ray photoelectron spectroscopy and surface X-ray scattering”. H.K. acknowledges the same LDRD grant for support for salary. The experiments were performed at beamline 11.0.2 of the Advanced Light Source, a U.S. DOE Office of Science User Facility, which is funded under contract no. DE-AC02-05CH11231. Q.L. acknowledges the ALS Postdoctoral Fellowship Program for the fellowship. H.P.M. has been supported for salary by the U.S. Department of Energy (DOE) under Contract No. DE-SC0014697. F.B. acknowledges funding through the LDRD project “Ultrafast Science Beyond Pump–Probe”. H.B. and M.B. also acknowledge support by the Director, Office of Science, Office of Basic Energy Sciences, and by the Division of Chemical Sciences, Geosciences and Biosciences of the U.S. Department of Energy at LBNL under Contract No. DE-AC02-05CH11231. Authors would like to thank ALS technical and other staff for their support, and acknowledge Caroline Bavasso for her contributions to the sample manipulation automation.

AIP PUBLISHING DATA SHARING POLICY

The data that support the findings of this study are available from the corresponding author upon reasonable request.

## AUTHOR INFORMATION

### Corresponding Author

* SNemsak@lbl.gov